\documentclass[useAMS,usenatbib]{mn2e}
\usepackage{graphicx}
\usepackage{natbib}
\usepackage{amssymb}
\usepackage{setspace}
\usepackage{amssymb}
\usepackage{epsfig,color}
\usepackage{multicol}

\long\def\symbolfootnote[#1]#2{\begingroup%
\def\thefootnote{\fnsymbol{footnote}}\footnote[#1]{#2}\endgroup} 

\title[Mass-density relationship in molecular cloud clumps]
  {Mass-density relationship in molecular cloud clumps}
\author[Donkov, Veltchev \& Klessen]
  {Sava Donkov$^{1\,\star}$, Todor V.~Veltchev$^{2,3}$, and Ralf S.~Klessen$^3$ \\
  $^1$Department of Applied Physics, Technical University, 8 Kliment Ohridski Blvd., 1000 Sofia, Bulgaria \\
  $^2$University of Sofia, Faculty of Physics, 5 James Bourchier Blvd., 1164 Sofia, Bulgaria\\
  $^3$Institute of Theoretical Astrophysics, Albert-\"Uberle-Str. 2, 69120 Heidelberg, Germany}

\date{Submitted 2011 Xxxxx XX}

\pagerange{\pageref{firstpage}--\pageref{lastpage}} \pubyear{2011}

\begin{document}

\label{firstpage}

\maketitle

\begin{abstract}
We study the mass-density relationship $n\propto m^x$ in molecular cloud condensations (clumps), considering various equipartition relations between their gravitational, kinetic, internal and magnetic energies. Clumps are described statistically, with a density distribution that reflects a lognormal probability density function (pdf) in turbulent cold interstellar medium. The clump mass-density exponent $x$ derived at different scales $L$ varies in most of the cases within the range $-2.5\lesssim x \lesssim-0.2$, with a pronounced scale dependence and in consistency with observations. When derived from the global size-mass relationship $m\propto l^{\gamma_{\rm glob}}$ for set of clumps, generated at all scales, the clump mass-density exponent has typical values $-3.0\lesssim x(\gamma_{\rm glob}) \lesssim-0.3$ that depend on the forms of energy, included in the equipartition relations and on the velocity scaling law whereas the description of clump geometry is important when magnetic energy is taken into account.
\end{abstract}


\section{Introduction} 
Observational evidence testifies that stars are born in molecular clouds (MCs) which sizes and mean densities vary in large ranges: $0.1-100~\rm pc$ and $1-10^5~\rm cm^{-3}$, respectively. Typical sites of star formation are dense regions in MCs \citep{Enoch_ea_07, Andre_ea_10}, often associated with young stellar objects \citep[e.g.,][]{JB_06}, wherein local gravitational instability leads to collapse and/or fragmentation and formation of protostellar cores. According to the modern paradigm in the star formation (SF) theory, such condensations result from shocks, generated by supersonic turbulent flows \citep[][{see \citealt[][for a review]{MK04}}]{Elme_93, KHM_00, PN02, Kr_ea_07, Dib_ea_08, Kr_ea_09}.
\symbolfootnote[0]{$\star$~E-mail: savadd@tu-sofia.bg}

Two main results from the theory of incompressible turbulence, developed by \citet{K41}, are the scaling law for the velocity dispersion and the kinetic energy spectrum. Prominent characteristics of turbulence in the cold interstellar medium (ISM) are compressibility and supersonic flows which lead to shocks and discontinuous density fluctuations. On the other hand, the hierarchy of scales and the cascade of the kinetic energy in them, predicted in the Kolmogorov's theory, are present in the cold ISM as well, although with different scaling relations. Observational evidence are ``Larson's first and second laws'' \citep{Lars_81}:
\begin{equation}
\label{Larson_1}
 u \propto L^\beta
\end{equation}
\begin{equation}
\label{Larson_2}
 \langle n \rangle \propto L^\alpha~,
\end{equation}
where $L$ is the spatial scale, $u$ is the rms velocity dispersion and $\langle n \rangle$ is the mean density. ``Larson's laws'' were derived for a wide range of structures: from $\sim0.1$~pc (`coherent cores', \citet{Good_ea_98}) up to $\sim100$~pc (clouds and cloud complexes), and revisited many times in observational and numerical works \citep{Solom_ea_87, MG_88, VS_ea_97, BMc_02, DBH_04, Hey_ea_09}. An issue that is still not clarified is whether the scaling exponents $\alpha$ and $\beta$ are interdependent -- e.g., considering compressible fluid in a statistical steady state \citep{Kr_ea_07}, or assuming equipartition of energies in MC clumps, obeying the above relations \citep{BP06}. On the other hand, a recent observational study of giant MCs \citep{Hey_ea_09} indicates that the coefficient in ``first Larson's law'' scales with their surface density, as not expected for purely turbulent objects but for ones wherein gravity plays a significant role \citep{BP_ea_11}. 

The structures considered in this work are condensations in MCs, generated by turbulent shocks at a given spatial scale $L$. Hereafter in this Paper, we label those condensations {\it clumps}, irrespectively of their density $n$ and its ratio to the mean scale density\footnote{By use of this ratio, some authors introduce subcategories like `voids' and `(dense) structures'.} $\langle n \rangle(L)$. A turbulent cascade is considered within a range of scales $30\gtrsim L \gtrsim 0.1$~pc, which corresponds to large regions within giant MCs and cloud complexes down to dense cloud cores. The upper limit is set to provide the molecular phase of the ISM and the applicability of an assumption of isothermality whereas it is probably the very process of cloud formation that drives the internal turbulence \citep[see][]{KH_10}. The chosen lower spatial limit is a typical transsonic scale at which the thermal energy per unit mass becomes comparable to the turbulent one. (For further details we refer the reader to the implementation of this approach in the model of the stellar initial mass function of \citealt{VKC_11}.) Our study is based essentially on the results of state-of-the-art 3D numerical simulations \citep{Dib_ea_07, Kr_ea_07, Shet_ea_10}. These simulations are able to reproduce the inertial range of turbulence only for small scales: $0.05\lesssim L \lesssim 1.5~\rm pc$. To enlarge the scope of our study, including clumps generated at scales up to $\sim30~\rm pc$, we refer as well to the classical 2D simulation of \citet{Pass_ea_95}, with length unit 1~kpc, and its implications for the energy balance of clouds and clumps \citep{BP_VS_95}.  

It is natural to expect a mass-density relationship for clumps formed in the cloud via turbulent fragmentation. This would set a physical link between them and protostars. Moreover, a power-law mass-size relationship for clouds and clumps in the ISM, termed ``Larson's third law'', is confirmed both from observations \citep{Kauf_ea_10, LAL_10} and numerical simulations (e.g., Shetty et al. 2010). Then, if ``Larson's second law'' (equation~\ref{Larson_2}) holds for small clumps and cores, although with a varying value of the exponent $\alpha$, a relationship between clump mass $m$ and clump density $n$ should be expected. The present work is an attempt for theoretical substantiation of a power-law relationship $n \propto m^x$. The exponent $x$ is estimated, assuming equipartition between various forms of energy within the clumps. 

In Section~\ref{nm_relationship} we present the physical assumptions on which our study is based, define the supposed clump mass-density relationship and explain the choice of its normalization. Our approach to estimate the power-law exponent $x$ as a function of the spatial scale $L$ is described in Section~\ref{approaches}. In Section~\ref{results} we present the results, compare them with observations of MCs and discuss the implications of different equipartition relations for the clump mass-size relationship. Section~\ref{Summary} summarizes the conclusions from this study and sketches its possible extensions.

\section{Definition of the clump mass-density relationship}   \label{nm_relationship}
\subsection{Basic physical assumptions and scaling laws}
\label{basic_physics}
Our consideration refers to an early stage of the MC evolution when turbulence is the key factor for structure formation at any spatial scale while gravity takes over at small scales. Driven turbulence at a given $L$ is fully developed at about one crossing time $t_{cr}(L)$ and saturated at $\sim1.5\,t_{cr}$ when a steady state is achieved \citep{Pass_ea_95, FKS08, Fed_ea_10}. At the latter epoch, gravity starts to play an essential role in the global evolution of the cloud \citep{Elme_00, VS_10} -- its total gravitational energy becomes comparable to the kinetic and internal energy \citep{VS_ea_07, ZAVS_11}. The fully saturated turbulence provides a wide inertial range of scales with well defined scaling laws for density, velocity and magnetic field. Dense structures in the initially diffuse interstellar medium form along the magnetic field  lines and become molecular, self-gravitating and magnetically supercritical at roughly the same timescale \citep{VS_ea_07, VS_10}, with densities of a few hundred $\rm cm^{-3}$ or more \citep{Crutch_ea_10}. Various simulations show \citep{Pass_ea_95, BP_VS_95, CB_05, Henne_ea_08, Baner_ea_09, Shet_ea_10, Dib_ea_10} that such condensations have typical sizes of MCs and clumps within them which fall into the inertial range of the natal environment. Also, the temporal stationarity within the inertial range allows for application of the ergodic hypothesis: a time averaged physical quantity can be replaced by its ensemble- or spatially averaged counterpart.

We assume also that the scaling laws do not depend on the turbulent forcing which is a good first-order approximation \citep{FKS08, Fed_ea_10}. The scaling of velocity is generally described by the ``Larson's first law'' (equation~\ref{Larson_1}) while the derived value of its exponent $\beta$ varies in relatively wide range. \citet{Pad_ea_06, Pad_ea_09} found $\beta=0.41-0.43$ from estimation of the turbulent power spectrum in MCs, while \citet{HB_04} obtained $\beta=0.65$  from observational structure functions of giant MCs and Monte Carlo modeling. Both works combine observational data with synthesized maps but apply different methods. A value $\beta\sim0.43$ was obtained also by \citet{Dib_ea_07}, from the correlation of velocity dispersion with size found in their simulations. In this Paper, we adopt the alternative values $\beta=0.42$ and $0.65$, which reflect the variation of this exponent and are close to its limit values. They are reproduced from the numerical simulations of \citet{Fed_ea_10} by use of the Fourier power spectrum and the PCA method, in case of pure solenoidal turbulent forcing.  

The density scaling law in the form of ``Larson's second law'' (equation~\ref{Larson_2}) is derived self-consistently at each scale, since it is interdependent with the assumed mass-density relationship for MC clumps (see Section \ref{clump_relationships} and \ref{approaches}). We suppose that it does {\it not} affect the scaling of the velocity. That is generally consistent with the very weak correlation of the local Mach number with density, found by \citet{Fed_ea_10}.

The dependence of the magnetic field $B$ on the spatial scale is obtained from its relation to the mean density. The latter is widely adopted to be $B\propto \langle n\rangle^{0.5}$, based on the study of \citet{Crutch_99}. However, as shown in \citet{Crutch_ea_10}, a refined analysis of the available data for MCs yields a steeper slope ($\sim2/3$), in consistency with the theoretical predictions for spherical collapse \citep{Mestel_66}. Therefore we use the form:
\begin{equation}
\label{B_scaling}
B = B_0 \Big(\frac{\langle n\rangle}{n_{\rm crit}}\Big)^{0.67}~,
\end{equation}
where $n_{\rm crit}$ is a critical density at which atomic clouds become self-gravitating, molecular and magnetically supercritical \citep{VS_10}. A pronounced scaling law $B(\langle n\rangle)$ exists for $\langle n\rangle\gtrsim n_{\rm crit}$, where we choose values of $n_{\rm crit}=150~\rm cm^{-3}$ (for molecular hydrogen) and $B_0=10~\rm \mu G$, that correspond to its upper envelope (see Fig. 1 and 5 in \citet{Crutch_ea_10}). 

The clouds and clumps in our consideration are assumed to be isothermal with typical temperature $T=10$~K. This is a good approximation for mean densities $\langle n \rangle \gtrsim 10^2$~cm$^{-3}$ \citep{Henne_ea_08} and corresponds to scales $L\lesssim 30$~pc \citep{Lars_81}.

\subsection{Clump density distribution and the concept of `average clump ensemble'}
The network of interacting shocks generated by supersonic turbulent flow causes density fluctuations in MCs. As demonstrated from many numerical simulations \citep{Kless_00, LKM03, P_ea07, Kr_ea_07, Dib_ea_08, FKS08, Fed_ea_10}, an appropriate statistical description of the distribution of the density $n$ per unit volume is a standard lognormal probability density function (pdf):
\begin{equation}
\label{eq_pdf}
p(s)\,ds=\frac{1}{\sqrt{2\pi \sigma^2}}\,\exp{\Bigg[-\frac{1}{2}\bigg( \frac{s -s_{\rm max}}{\sigma}\bigg)^2 \Bigg]}\,ds~,
\end{equation}
with standard deviation (stddev) $\sigma$ which depends on the Mach number $\cal M$ for supersonic flows and determines the peak position: 
\begin{equation}
\label{max_PDF}
s_{\rm max}=-\frac{\sigma^2}{2}~,
\end{equation}
\begin{equation}
\label{sigma_PDF}
\sigma^2={\rm ln}\,(1+b^2\,{\cal M}^2)~,
\end{equation}
where $s\equiv \ln(n/\langle n \rangle)$ and $\langle n \rangle$ is the mean density in the considered volume which we estimate by use of equation~(\ref{Larson_2}). The quantity $b$ is usually called {\it turbulence forcing parameter} and varies in the range $0.2-1.0$, depending on the driving type \citep{Kr_ea_07, FKS08, Fed_ea_10}. In the present work we vary the values $0.30\le b\le 0.70$, depending on the approach and the data extrapolation. Such variations of the parameter $b$ affect the velocity scaling slopes in the range $0.42-0.48$ \citep{Fed_ea_10}, much less than the one covered in our study. 

The pdf gives a distribution of density contrasts generated through turbulent velocity field. How can one relate a density contrast to a specific spatial object with its physical parameters (size, mass etc.) like a clump in the ISM? Our study is focused not on a set of individual clumps with their characteristics but on a group of statistical objects, labeled {\it average clump ensemble}, representative for all clumps generated within a considered volume $L^3$. The average clump ensemble at given scale $L$ results from ensemble averaging over the variety of Galactic clouds and cloud complexes.  

The peak of the pdf (equation~\ref{max_PDF}) corresponds to the most probable member of this ensemble (`typical clump') with density:
\begin{equation}
\label{eq_nc}
 n_c=\langle n \rangle\,\exp(s_{\rm max})=\langle n \rangle\exp\Big(-\frac{\sigma^2}{2} \Big)
\end{equation}
Hereafter, we denote the parameters of the `typical clump' by subscript $\rm c$. The relation (\ref{eq_nc}) includes dependencies on the spatial scale $L$ from equations \ref{Larson_1} (through the Mach number dependence of the stddev, Eq.~\ref{sigma_PDF}) and \ref{Larson_2}. The range of densities in the average clump ensemble is defined as $\pm\sigma/2$ from the most probable density $n_c$. That yields variations in respect to the mean scale density $\langle n \rangle$:
\begin{equation}
\label{ens_limits}
 \langle n \rangle\,\exp\Big(s_{\rm max}-\frac{\sigma}{2}\Big)\le n \le \langle n \rangle\,\exp\Big(s_{\rm max}+\frac{\sigma}{2}\Big)
\end{equation}
The density of a `typical clump' differs from $\langle n \rangle$ within an order of magnitude for all considered scales $\lesssim30$~pc. As will be demonstrated in Section \ref{results}, its size $l_c$ is substantially less than the scale $L$. 

\subsection{Clump relationships and their normalization}
\label{clump_relationships}
As mentioned in Section 1, we assume a power-law clump mass-density relationship:
\begin{equation}
\label{eq_n-m}
{\rm ln}\Big(\frac{n}{n_0}\Big)=x\,{\rm ln}\Big(\frac{m}{m_0}\Big)
\end{equation}
where $n_0$ and $m_0$ are units of normalization. Recalling that the average clump ensemble and the `typical clump' are statistical objects, equation~(\ref{eq_n-m}) implies a clump size-density relationship: 
\begin{equation}
\label{eq_n-l}
{\rm ln}\Big(\frac{n}{n_0}\Big)=\alpha_x\,{\rm ln}\Big(\frac{l}{l_0}\Big)~,~~~\alpha_x=\frac{3x}{1-x}~,
\end{equation}

We denote the clump density scaling exponent at a given scale by $\alpha_x$ with reference to the mean density scaling exponent $\alpha$ (equation \ref{Larson_2}). Note that these two quantities are not necessarily equal - this would be the case if the clumps, generated at some scale $L$, obey the scaling law of the mean density. Further in Section \ref{clump_energies} we assume a self-consistent density scaling, requiring $\alpha_x\equiv \alpha$ for each $L$.
 
Using equations \ref{eq_n-m} and \ref{eq_n-l}, characteristic mass $m_c$ and size $l_c$ are ascribed to the `typical clump' with density $n_c$:
\begin{equation}
\label{ml_clumps}
 m_c=m_0\Big(\frac{n_c}{n_0} \Big)^{\frac{1}{x}}
\end{equation}
\begin{equation}
\label{nl_clumps}
 l_c=l_0\Big(\frac{n_c}{n_0} \Big)^{\frac{1-x}{3x}}
\end{equation}

By use of equations~(\ref{eq_pdf}), (\ref{eq_n-m}) and (\ref{eq_n-l}), one derives lognormal mass and size distributions with positions of the maximums and widths as follows:
\begin{eqnarray}
\label{param_x}
s_{\rm max,\,m}=s_{\rm max}/x~,~~~~\sigma_m=\sigma/|x| \\
s_{\rm max,\,l}=s_{\rm max}/\alpha_x~,~~~~\sigma_l=\sigma/|\alpha_x|
\end{eqnarray}

A natural choice of the normalization unit $n_0$ is the mean density at the scale in consideration $L$: $n_0\equiv\langle n \rangle$. The choice of the size normalization unit $l_0$ should be made taking into account typical relative sizes of clumps in respect to the scale in consideration. It is appropriate to choose $l_0$ to be small and proportional to the scale size $L$:
\begin{equation}
\label{kappa_eq}
 l_0=\kappa L~~,
\end{equation}
where the dimensionless parameter $\kappa$ could be interpreted as mapping resolution of the scale volume. Its appropriate value should be chosen to be of order of $10^{-2}$ to achieve a distinction of  substructures, significantly smaller than the spatial scale $L$ and significantly larger than the scale of dissipation. Further comment on that issue follows in Sect.~\ref{M-R_relations}. 
 
The relation between the normalization units $n_0$, $m_0$ and $l_0$ is obtained from the requirements for volume and mass conservation, as shown in the Appendix \ref{appendix_a}. In view of our simple definition of a clump as a condensation of arbitrary density (see Section 1), a scale volume is considered to be totally occupied by $N$ clumps. The requirement for volume conservation then yields:
\begin{equation}
\label{volume_conservation}
 L^3\simeq Nl_0^3 \exp\Big(\sigma^2\times\frac{(1-x)(1-2x)}{2x^2}\Big)
\end{equation}
 with the obvious relation between $\kappa$ and $N$ (cf. equation \ref{kappa_eq}):
\begin{equation} 
\label{N_clumps}
 N=\frac{1}{\kappa^3}\exp\Big(\sigma^2\times\frac{(x-1)(1-2x)}{2x^2}\Big) 
\end{equation}
 
On the other hand, the total mass of the scale is $M=a\langle \rho \rangle L^3=a\mu\langle n \rangle L^3=a\rho_0 L^3$ where we adopt $\mu=2.4m_p$ and $a$ is a dimensionless parameter of order unity which reflects the clumps geometry. The requirement for mass conservation leads to:
\begin{equation}
\label{mass_conservation}
 M\simeq Nm_0\exp\Big(\sigma^2\times\frac{1-x}{2x^2}\Big)
\end{equation}

Then, from equations \ref{volume_conservation} and \ref{mass_conservation}, one obtains the relation between the normalization units:
\begin{equation}
\label{norm_units}
 a\frac{\rho_0 l_0^3}{m_0}=\exp\Big(\sigma^2\times\frac{1-x}{x}\Big)
\end{equation}

In this work, we assume spherical shapes of clumps and spatial scales. Thus $a \equiv \pi/6$, but the result is principally the same when cubic shapes are considered. 

\section{Approach to estimate the mass-density exponent}   \label{approaches}

The power-law exponent $x$ in equation~(\ref{eq_n-m}) depends on the clump physics. It was adopted as a free parameter in the model of the initial stellar mass function of \citet{VKC_11} without theoretical substantiation. In this paper we use different approaches to estimate $x$, based on equipartition relations between the gravitational, kinetic, internal and magnetic energy of the MC clumps.  

\subsection{Equipartitions of energies} \label{energy_equipartitions}
The energy balance in an object of arbitrary form in the ISM is determined by use of the virial theorem (VT). Its Eulerian form is given elsewhere \citep{MZ_92, BP06, Dib_ea_07}. In the current work we study the effect of relations only between the VT volume energy terms since they could be estimated more easily from observations. The gravitational, kinetic (turbulent) energy, thermal (internal) energy and magnetic energy per unit volume are defined as:
\[ W=-\frac{1}{2V}\int \rho\phi\,dV~,~~E_{\rm kin}=\frac{1}{2V}\int\limits_V \rho u^2\,dV,\]
\[ E_{\rm th}=\frac{3}{2V}\int\limits_V P\,dV,~~~E_{\rm mag}=\frac{1}{8\pi V}\int\limits_V B^2\,dV \]

where $P$ is the pressure, $\phi$ is the gravitational potential and $V$ is some arbitrary volume. The volume normalization is appropriate for clumps since their main parameters are derived (Section \ref{clump_relationships}) by using a volume weighted pdf (equation \ref{eq_pdf}). For better readability, we use hereafter simply the term `energy' instead of `energy per unit volume' when referring to the quantities listed above. 

There is no requirement for virial equilibrium in our treatment. \citet{BP06} showed that such an assumption is too strong; it is also not consistent both with observations and numerical simulations of the ISM \citep{Shad_ea_04, Dib_ea_07, Dib_Kim_07}. In fact, MCs seem to be in {\it energy equipartition}. In our approach we assume relations between $W$, $E_{\rm kin}$, $E_{\rm th}$ and $E_{\rm mag}$ of the clumps that hold on statistical terms, for the average clump ensemble. These relations are labeled `equipartitions' for simplicity. 

Gravity and turbulence are the main factors for clump dynamics at the considered stage of MC evolution (Section \ref{basic_physics}) -- developed turbulence, inherited from the earlier epoch of cloud formation \citep{KH_10}, shapes the density structure while gravity tends to dominate at small scales, in local condensations. Therefore the basic statistical equipartition for clumps should be a functional dependence like:
\begin{equation}
\label{grav_kin}  
 |W|\sim f_{\rm gk}E_{\rm kin}~,
\end{equation}
where $f_{\rm gk}$ is a coefficient of proportionality. For self-gravitating clumps of larger sizes (clouds), a `virial-like' relation $f_{\rm gk}\sim1-2$ could be expected \citep{VS_ea_07} while for clumps of sizes typical for dense cloud cores, $\lesssim0.1-0.2$~pc, a more relevant equipartition should include the internal energy as well. In order to study different stages of clump evolution when gravity takes gradually over, we adopt further fiducial values $1\le f_{\rm gk}\le 4$.

Various equipartitions were found by \citet{BP_VS_95} (hereafter, BV95) from analysis of 2D MHD simulations \citep{Pass_ea_95}. These authors delineated clumps and clouds within a large range of sizes $1\lesssim l \lesssim 100~\rm pc$ (see their Fig. 2) through isodensity contours at densities 4 to 16 times larger than the mean density in the simulational box ($L=1$~kpc). Some relations BV95 discovered are:
\begin{itemize}
\item Equipartitions of the kinetic energy vs. magnetic and thermal energy (cf. Fig. 4 in BV95), respectively:
\begin{equation}
\label{kin_th}  
 E_{\rm kin}\sim E_{\rm th}
\end{equation}
\begin{equation}
\label{kin_mag}  
 2E_{\rm kin}\sim E_{\rm mag}
\end{equation}
 \item Equipartition of the gravitational vs. kinetic and magnetic energy:
 \begin{equation}
  \label{noth_mag}  
  |W|\sim 2E_{\rm kin}+E_{\rm mag}
 \end{equation}
\end{itemize}
In the work of BV95, the energy of gravity due to the ambient medium was not taken into account in the term $|W|$. We allow for moderate gravitational influence from the external cloud in our consideration of the typical clump (Section \ref{clump_energies}), neglecting possible interaction with the large-scale environment. 

The tendency towards equipartition among different forms of energy found by BV95 stimulated us to include additional equipartitions of the gravitational energy in our study:
\begin{equation}
\label{grav_th}  
 |W|\sim f_{\rm gt} E_{\rm th}
\end{equation}
\begin{equation}
\label{grav_mag}  
 |W|\sim f_{\rm gm} E_{\rm mag}
\end{equation}
with coefficients of proportionality $f_{\rm gt}$ and $f_{\rm gm}$. 

The relations (\ref{grav_kin}) - (\ref{grav_mag}) shall be considered as a tool to estimate the contribution of differenf forms of energy in the clump physics. Whether their use is justified or not is to be decided on the base of comparison between our results (Section~\ref{results}) and the referred observational and numerical evidence.

\begin{figure*} 
\begin{center}
\includegraphics[width=1.\textwidth]{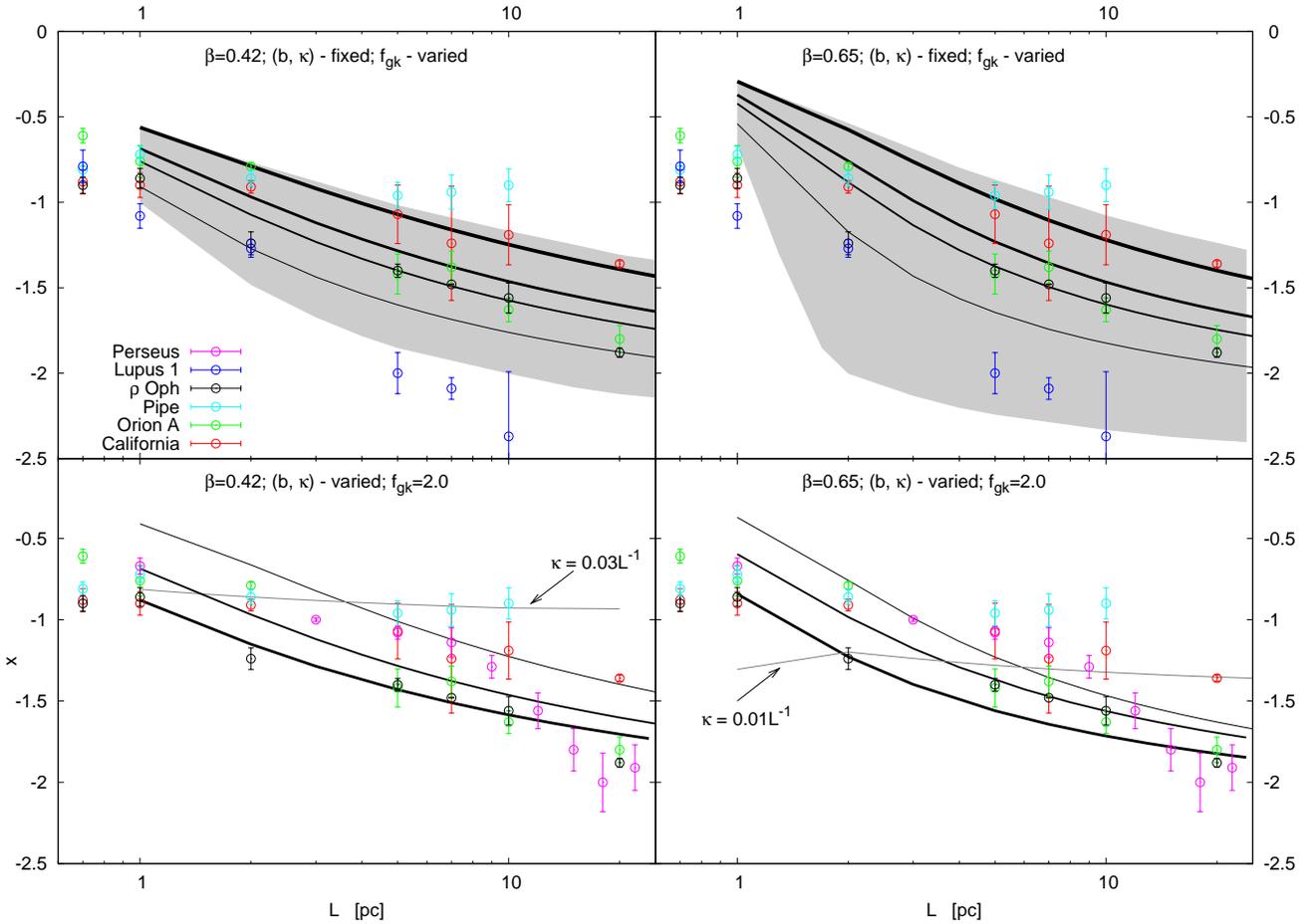}
\vspace{0.6cm}
\caption{Scale dependence of the mass-density exponent $x$, derived from equipartitions of gravitational vs. kinetic energy (lines; equation~\ref{grav_kin}) for the 'typical clump' and adopting alternative values of the velocity scaling exponent. {\it Top:} fixed forcing parameter and mapping resolution (left: $b=0.50$, $\kappa=0.014$; right: $b=0.30$, $\kappa=0.01$) and varied $f_{\rm gk}=1.0,\,1.5,\,2.0,\,4.0$ (increasing linewidth). The shaded areas denote the zone of solutions for the average clump ensembles.  {\it Bottom:} strongly self-gravitating clumps ($f_{\rm gk}=2.0$) and varying forcing parameter $b=0.30,\,0.50,\,0.70$ (increasing linewidth), simultaneously with $0.03\ge\kappa\ge0.001$ . A typical solution for a scale-dependent $\kappa$ is drawn (grey lines; see text for details). Observational estimates $x(\gamma)$ according to \citet{LAL_10} for some clouds and cloud complexes (open circles) are plotted for comparison.}
\label{fig_gr_kin}
\end{center}
\end{figure*}

\subsection{Clump energies and equipartitions functions}
\label{clump_energies}
The equipartition equations from the previous Section can be written for clumps generated at a given scale within the inertial range, by use of the scaling laws of density, velocity and magnetic field strength (Section~\ref{basic_physics}). The energies per unit volume are calculated for the `typical clump' while the density range of the `average clump ensemble' (equation~\ref{ens_limits}) defines their variations. Since the `typical clump' density $\rho_c=\mu n_c$ is proportional (equation~\ref{eq_nc}) to the mean scale density $\langle \rho \rangle(L)\propto L^{\alpha}$ ($\alpha<0$), this approach is comparable with the one of BV95 who delineate clumps imposing density thresholds, indicative for the corresponding scale (see their Fig. 2).

One gets for the gravitational energy:
\begin{equation}
\label{W_clumps}
 |W|=z_c\,\frac{3}{5}G\frac{m_c}{l_c/2}\rho_c
\end{equation}
where the coefficient $z_c$ accounts for the contribution of the mass outside the clump to its gravitational energy. As demonstrated by \citet{BP_ea_09}, characteristic values of $z_c$ vary between 1 (vanishing gravitational influence) and 2 (strong gravitational influence from the external cloud). That range is applicable for scales below the sizes of giant MC, like in our study. We adopt $z_c=1.5$ in all considered cases. 

The volume term of the internal energy is obtained straightforwardly in view of the assumed spherical symmetry:
\begin{equation}
\label{ei_clump}
 E_{\rm th}=\frac{3}{2}\frac{\Re}{\mu}\rho_c T~,
\end{equation} 
where $\Re$ is the gas constant. 

The density of the `typical clump' $\rho_c$ scales with its size according to equation~(\ref{nl_clumps}). If formation of clumps is a turbulent phenomenon, this clump size-density relationship should be a self-similar extension of the scaling of density according to the ``Larson's second law''\footnote{~Indeed, \citet{Lars_81} did not make a sharp distinction between clouds and clumps (structures within the clouds) in the sample he studied.}(equation~\ref{Larson_2}). Hence, one would expect that the mean density scaling exponent $\alpha$ obeys the relation
\begin{equation}
 \label{dens_scaling}
 \alpha=\alpha_x=\frac{3x}{1-x}~.
\end{equation}
We derive $\alpha$ and $x$ at given scale $L$ in a self-consistent way as follows. We start with some fiducial value $\alpha=\alpha^{(0)}$ and derive a solution $x=x^{(0)}$ which yields through equation~(\ref{dens_scaling}) a new value $\alpha=\alpha^{(1)}$. Then a new value $x=x^{(1)}$ is calculated which again is used to calculate $\alpha$ and so forth. The iterative procedure converges fast and leads to self-consistent values of $\alpha$ and $x$.

The magnetic energy could be also expressed in terms of $\rho_c$. In contrast to the treatment of clump density demonstrated above, we do not assume a self-similar scaling of $B$. The general magnetic field scaling $B(L)$ is given by equation~\ref{B_scaling}. On the other hand, since most of the clumps in a given `average ensemble' at scale $L$ are underdense in relation to the scale mean density $\langle n \rangle$ (equation \ref{ens_limits}), the magnetic field $B(l)$ within the ensemble is asummed to obey a scaling law of shallower slope $1/2$ \citep{Crutch_99}. This combination of different scaling laws of $B$ provides the maximal possible values of $E_{\rm mag}$ in the corresponding equipartition equations. Thus the expression for magnetic energy of a `typical clump' reads:
\begin{equation}
\label{em_clump}
 E_{\rm mag}=\frac{B^2(L)}{8\pi}\frac{n_c}{\langle n \rangle}~,
\end{equation} 
In fact, such description is appropriate for objects with higher energies \citep{Crutch_ea_10}.

Lastly, to calculate the kinetic energy of the `typical clump', one should implement also the velocity scaling law:
\begin{equation}
\label{ek_clump}
 E_{\rm kin}=\frac{1}{2}\rho_c u_c^2=\frac{1}{2}\rho_c\,u_0^2 \bigg(\frac{l_c}{4~{\rm pc}}\bigg)^{2\beta},~~~u_0=2~{\rm km/s}
\end{equation}
where the chosen normalization units of size and velocity are based on the simulations of \citet{Dib_ea_07}.

Now the exponent $x$ of the clump mass-density relationship is derived as a zero of the ``equipartition functions'', that correspond to the relations in Section~\ref{energy_equipartitions}. The way to obtain these functions and their form (equations~\ref{equipf_grav_kin}-\ref{equipf_grav_mag}) are given in the Appendix \ref{appendix_b}. 

\begin{figure*} 
\begin{center}
\includegraphics[width=1.\textwidth]{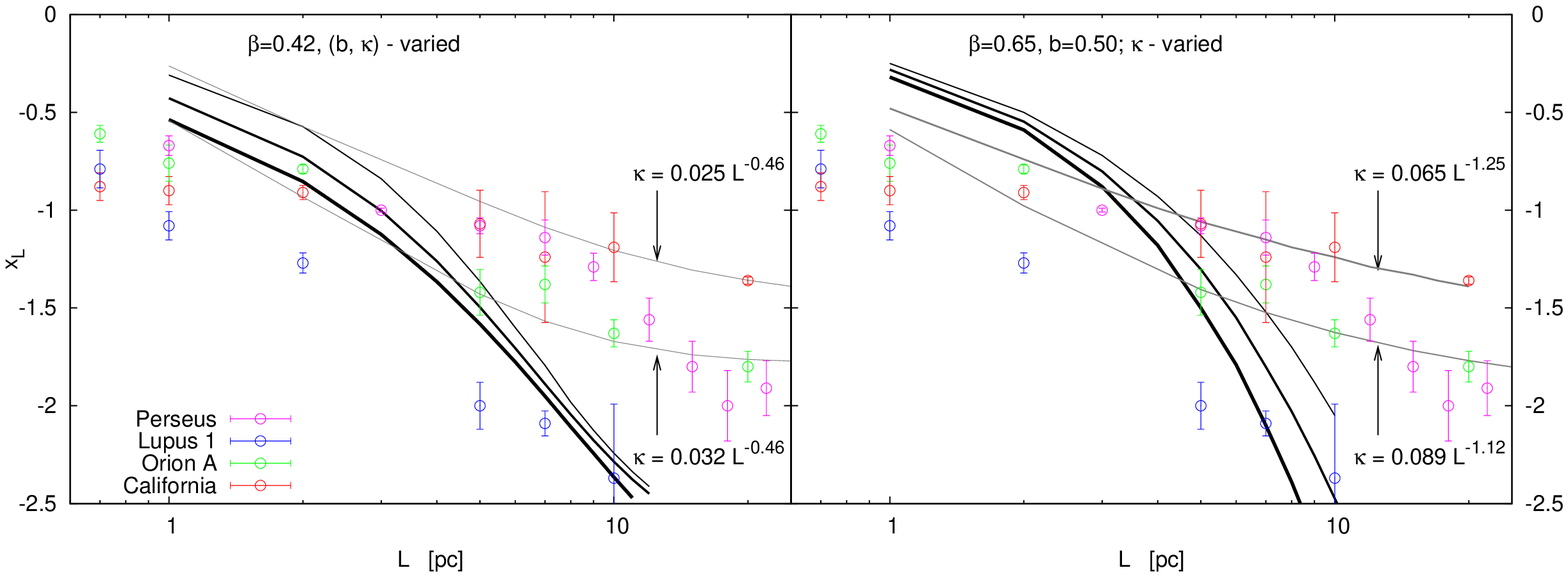}
\vspace{0.6cm} 
\caption{The mass-density exponent $x$, derived from equipartition of the gravitational vs. kinetic and magnetic energy (lines; equation~\ref{noth_mag}) for the 'typical clump' and adopting alternative values of the velocity scaling exponent. {\it Left:} Varied forcing parameter $b=0.30, 0.50, 0.70$ (black lines, increasing linewidth), simultaneously with $\kappa=0.035, 0.017, 0.010$, respectively; {\it Right:} Fixed forcing parameter $b=0.50$, varying only the mapping resolution (black lines, $0.01\le\kappa\le0.016$). Other designations are the same like in Fig.~\ref{fig_gr_kin}.}
\label{fig_noth_pan}
\end{center}
\end{figure*}

\section{Results and discussion}	\label{results}
\subsection{Mass-density exponent at a given scale}
The exponent $x$ of the clump mass-density relationship derived at a given scale originates from the density distribution of the `average clump ensemble' (equation~\ref{ens_limits}) and, hence, from the adopted scaling laws of density and velocity\footnote{~Due to the scale dependence of $\sigma$ through the Mach number, cf. equation~\ref{sigma_PDF}.}. Thus $x$ is essentially a {\it parameter of the scale}. An observational counterpart of this quantity could be derived from mass-size relations for (usually) nested structures within MCs, obtained from extinction maps (e.g. Lombardi, Alves \& Lada 2010; hereafter, LAL10) or column density maps (e.g. Kauffmann et al. 2010):
\begin{equation}
\label{mr_relation}
 M_{\rm s} \propto R_{\rm s}^{\gamma}~,
\end{equation}
where $M_{\rm s}$ and $R_{\rm s}=\sqrt{S/\pi}$ are the mass and the effective radius of a structure with area $S$. Note that in both mentioned observational studies, the mass-size exponent $\gamma$ is found to vary with $R_{\rm s}$. Considering such structures as scales of clump formation, i.e. $R_{\rm s}\sim L$, and taking into account the assumed self-consistent density scaling (equation~\ref{dens_scaling}), one derives from equation~(\ref{ml_clumps}) and (\ref{nl_clumps}):
\begin{equation}
\label{gamma_x_local}
 x=\frac{\gamma-3}{\gamma}
\end{equation}
Hence one is enabled to compare observational estimates of $x (\gamma)$ (relation \ref{gamma_x_local}) with the values of $x$ obtained by use of our approach from various equipartitions (equations \ref{grav_kin} - \ref{grav_mag}), at different scales $L$ considered as substructure entities of given MC. We chose for comparison the work of LAL10 since they studied the internal structure of a rich sample of clouds, covering a large range of scales ($0.01\lesssim R_{\rm s}\lesssim70$~pc). The latter partially coincides with range of cloud/clump sizes in the work of BV95 which inspired our equipartition approach.

Our method is statistical and, in this sense, universal for the variety of Galactic MCs. On the other hand, the results depend on several varying parameters: the forcing parameter $0.30\le b \le 0.70$, the velocity scaling exponent $0.42\le\beta\le0.65$ and the mapping resolution parameter $\kappa \lesssim 0.1$. The assumption for self-similar density and velocity scaling of the average clump ensemble sets a lower limit of the spatial scale: the `typical clump' size is required to fall within the inertial range, i.e. its minimal value must be $\sim 0.1$~pc. Since $l_c$ is typically about an order of magnitude less than the scale of generation for all equipartition methods, relevant results could be obtained for $L\gtrsim1$~pc.

\begin{figure} 
\begin{center}
\includegraphics[width=80mm]{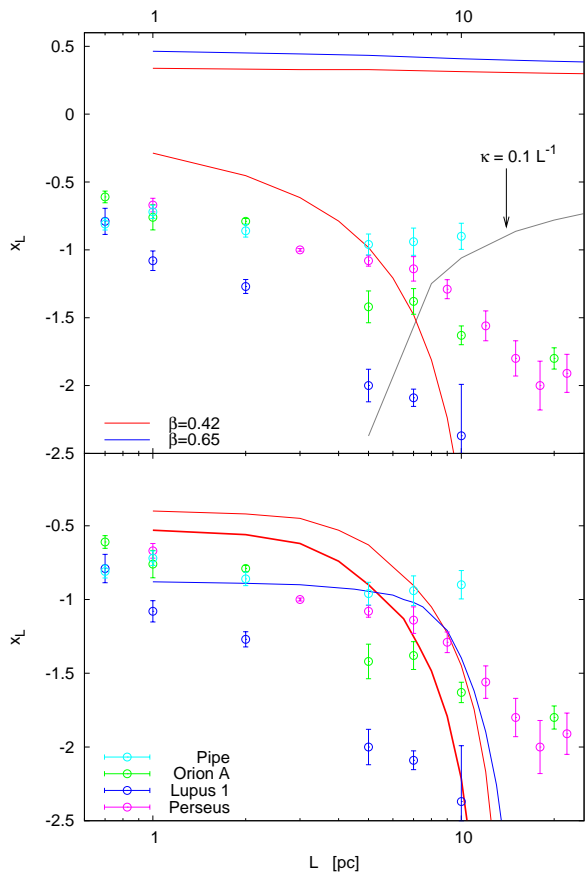}
\vspace{0.6cm} 
\caption{The mass-density exponent $x$, derived for equipartitions of the kinetic energy (lines; equations \ref{kin_th} and \ref{kin_mag}). {\it Top:} $E_{\rm kin}\sim E_{\rm th}$: effects of constant ($b=0.60$, $\kappa=0.024$, red line) vs. scale-dependent mapping resolution ($b=0.33$, grey line) are juxtaposed. Note the obtained positive solutions $x\approx \rm const > 0$ ($b=0.33,~\kappa=0.1$); {\it Bottom:} $2 E_{\rm kin}\sim E_{\rm mag}$: only solutions with a scale-dependent mapping resolution are possible ($\kappa \propto L^{-1}$). The effect of varying the forcing parameter $b=0.30, 0.50$ are illustrated (increasing linewidth).}
\label{fig_kinthmag}
\end{center}
\end{figure}

\begin{figure*} 
\begin{center}
\includegraphics[width=1.\textwidth]{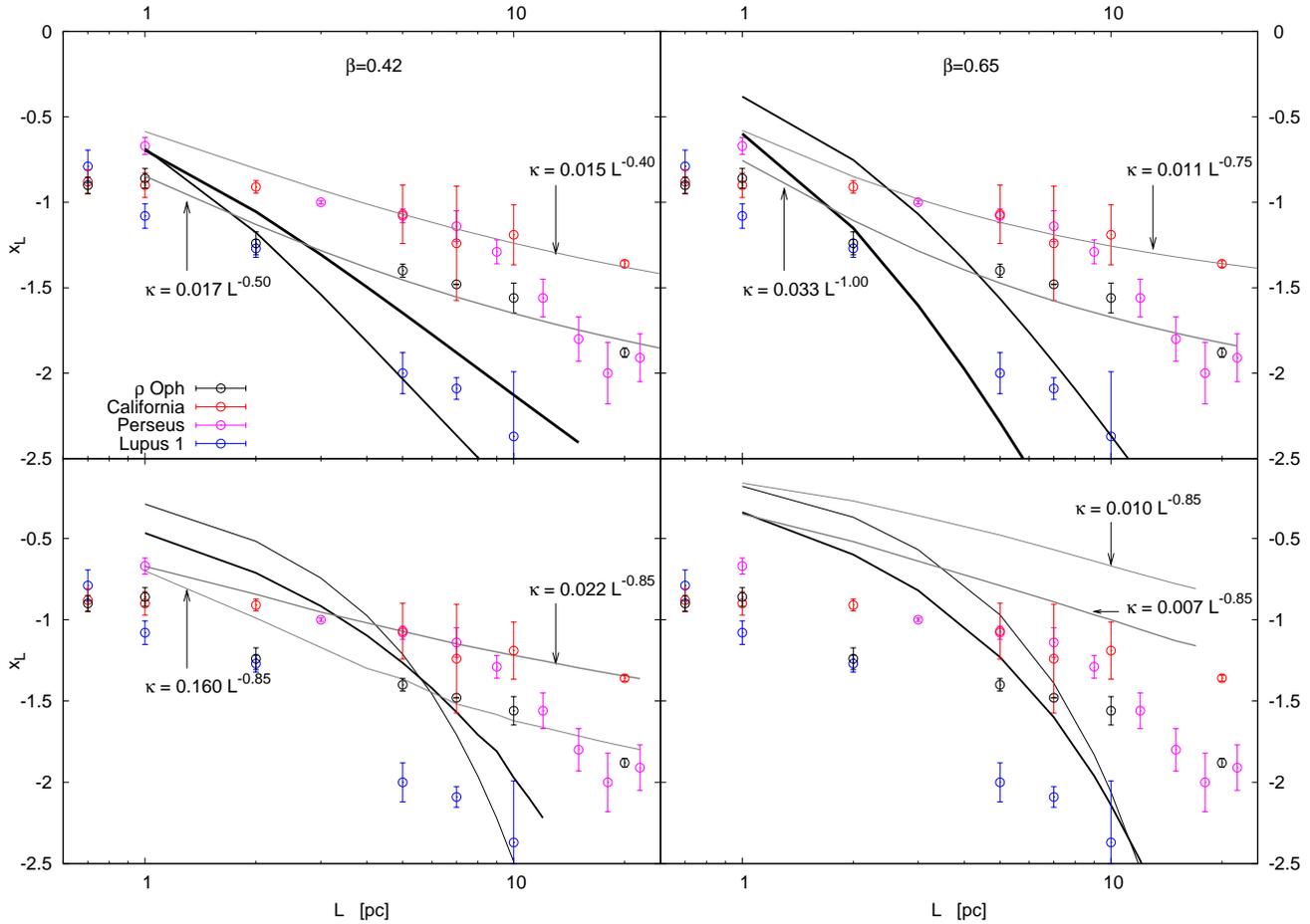}
\vspace{0.6cm} 
\caption{The mass-density exponent $x$, derived from equipartition of the gravitational vs. internal (equation~\ref{grav_th}) and magnetic energy (equation~\ref{grav_mag}). {\it Top} ($|W|\sim E_{\rm th},~f_{\rm gt}=1$): Solutions for $b=0.50, 0.70$ (increasing linewidth) for constant (black lines; $\kappa=0.023, 0.010$ /left/, $\kappa=0.013, 0.007$ /right/, respectively) and scale-dependent mapping resolution parameter (grey lines) are plotted; {\it Bottom} ($|W|\sim E_{\rm mag},~f_{\rm gm}=1$): solutions for $b=0.30, 0.70$ (increasing linewidth) for constant $\kappa$ (black lines; $\kappa=0.030, 0.008$ /left/, $\kappa=0.017, 0.006$ /right/, respectively). }
\label{fig_gr_magth_pan}
\end{center}
\end{figure*}

As evident from Fig.~\ref{fig_gr_kin}-\ref{fig_gr_magth_pan}, all approaches exhibit a general trend in three aspects: 
\begin{itemize}
 \item Negative values of $x$ are obtained independently of the approach. Their range at all considered scales is in agreement with the estimates from the most recent observations (Kauffmann et al. 2010, LAL10). Only the method $E_{\rm kin}\sim E_{\rm th}$ yields positive $x$ as well, clearly outside the zone of observational data (Fig.~\ref{fig_kinthmag}).
 \item Negative values of $x$ can vary from $-0.2$ down to $-2.5$ while the specific functional behavior (convexity/concavity, steepness) depends on the chosen set of values of the parameters $b$, $\beta$ and $\kappa$. When $(b, \beta)$ are fixed, increase of $\kappa$ leads to a shift of the curve $x(L)$ downwards (i.e. decrease of $x$), retaining its shape. On the other hand, increase of $b$ or $\beta$ steepens the curve at larger scales. Positive $x$ (obtained in the case $E_{\rm kin}\sim E_{\rm th}$) are virtually independent on $L$ and $b$; the curves $x(L)$ are slightly shifted vertically by variations $\kappa$ and $\beta$. Choice of a scale-dependent $\kappa$ generates shallow curves $x(L)$, in agreement with the structure of some larger cloud complexes.
 \item The coefficients of proportionality in equations \ref{grav_kin}, \ref{grav_th} and \ref{grav_mag} affect the curves $x(L)$ in different ways: variations of $f_{\rm gk}$ lead to vertical shift; the steepness of the curves is highly sensitive on variations of $f_{\rm gt}$; while variations of $f_{\rm gm}$ do not influence significantly the functional behaviour.
\end{itemize}

We sorted out the physically meaningful solutions for $x$ by {\it two criteria}: i) comparison with the observational data $x(\gamma)$, and ii) comparison with the expected range of the basic clump parameters from simulations like size, mass, density and mean density of the scale. Both criteria were applied simultaneously. 

The results obtained from the equipartition of the clump gravitational vs. kinetic (turbulent) energy fit well with some of the LAL10 data (Fig.~\ref{fig_gr_kin}), independent on the chosen velocity scaling. Variation of the factor $f_{\rm gk}$ (top panels), which may correspond to different evolutionary stages of a clump, allows for fitting the observed structure of some large cloud complexes like California, Orion A and $\rho$~Ophiuchi, while it is problematic to attain agreement with a `shallow structure' in terms of $x$ (i.e. $\gamma \approx \rm const$; Pipe nebula) or with a too `steep structure' (Lupus 1). The results are similar for strongly self-gravitating clumps ($f_{\rm gk}=2.0$, Fig.~\ref{fig_gr_kin}, bottom panels) when the turbulent forcing parameter is varied. The effect from variation of the mapping resolution parameter $\kappa$ (not illustrated) is a moderate steepening of the curves $x(L)$ at larger scales without essential change of their shape. Some additional opportunity to fit shallowly structured clouds like Pipe nebula gives the choice of a scale-dependent mapping resolution parameter (Fig.~\ref{fig_gr_kin}, bottom) $\kappa\propto L^{-a}$, where $0.4\lesssim a \lesssim 1.3$ for all equipartitions. That is a reasonable approach since the wide density and mass distributions produced at large scales (cf. equations~\ref{Larson_1} and \ref{sigma_PDF}) require a better mapping resolution (decrease of $\kappa$) in order to trace the substructures. However, such an approach needs refinement, taking into account the diversity of clump shapes.

In view of the range of possible values of $f_{\rm gk}$ and $b$, general agreement between the cloud structure, predicted from the equipartitions $|W|\sim f_{\rm gk}E_{\rm kin}$, and the LAL10 data for some clouds is not enough to determine the energy balance of the `typical clumps' in them. Rather, we aim to demonstrate the capacity of our statistical approach, combined with the equipartition equation to derive $x$. To model the cloud structure in more detail, one needs additional data about: i) the evolutionary status of the cloud which restricts the applicability of our concept (Sect.~\ref{basic_physics}); ii) the physical conditions in the cloud that determine and/or could be implied from its clump mass function. The latter is to be demonstrated in a subsequent paper.

The addition of a magnetic term in the equipartition of the gravitational vs. kinetic energy (equation~\ref{noth_mag}) for strongly self-gravitating clumps ($f_{\rm gk}=2$) and $\kappa=\rm const$ generates curves $x=x(L)$ with a different shape which cannot fit the observational data (Fig. \ref{fig_noth_pan}). On the other hand, adopting a scale-dependent $\kappa$ as mentioned above yields solutions that are consistent with the structure of the large cloud complexes; in contrast to the result when a magnetic term is missing (Fig.~\ref{fig_gr_kin}, bottom). That points to the importance of clump geometry when the magnetic field is considered\footnote{~Note, however, that in the assumed evolutionary time frame of our study clump geometry is determined mainly by gravity.}. 

Equipartitions of the kinetic energy (equations~\ref{kin_th} and \ref{kin_mag}) yield solutions with anomalous behavior of $x$; i.e. either lead to increasingly steeper slope or to extremely small values at small scales (Fig.~\ref{fig_kinthmag}, top). Equipartition $E_{\rm kin} \sim E_{\rm th}$ produces a family of curves with {\it positive} $x$, practically independent on scale. However, those solutions fail to meet the criteria we adopted: they lie outside the observational range and correspond to an unrealistic set of clump parameters, as will be shown below. Negative $x$ that lie in the domain of the LAL10 data are obtained in a very limited range of spatial scales $5\lesssim L\lesssim 10$~pc, but are clearly unrealistic. The results for the case $2 E_{\rm kin} \sim E_{\rm mag}$ are similar in terms of application (Fig.~\ref{fig_kinthmag}, bottom). Evidently, the clump mass-density relationship cannot be explained only by use of equipartitions of the kinetic energy, without invoking gravity. That is consistent both with time frame of our study: the transition between the turbulent epoch and the stage of hierarchical gravitational collapse, and with the supposed spherical symmetry of the clumps. Turbulence is the main factor for the formation of clumps but gravity is crucial to determine their fundamental physical characteristics. The latter suggestion is supported by the results obtained from equipartitions of the gravitational vs. thermal and vs. magnetic energy (equations~\ref{grav_th} and \ref{grav_mag}; Fig.~\ref{fig_gr_magth_pan}). The agreement with observations of large cloud complexes (California, $\rho$ Oph) is remarkable for choice of a scale-dependent $\kappa\propto L^{-a}$, except in the case $|W|\sim E_{\rm mag},~\beta=0.65$. The choice of constant $\kappa$ is able to explain `steeply-structured' clouds like Lupus 1, only by use of the classical Jeans approach, i.e. gravity vs. internal energy. The solutions for $f_{\rm gt}=2$ in equation~\ref{grav_th} are highly sensitive to the choice of $\beta$ and $\kappa$, retaining the shape of the curves, while those for $f_{\rm gm}=2$ in equation~\ref{grav_mag} practically coincide with those for $f_{\rm gm}=1$. (Therefore these cases are not illustrated in Fig.~\ref{fig_gr_magth_pan}.)

From the point of view of the current understanding of MC physics, we suppose that the better agreement of the results for the equipartitions of $W$ with observations of LAL10 could be explained with the evolutionary status of the clouds in which clumps are generated. \citet{BP_ea_11} suggest that the MCs undergo two stages: an epoch of hydrodynamic turbulence, followed by an epoch of a hierarchical gravitational collapse. The clumps considered in this work are typical objects existing in the transitional epoch between them. 

\subsection{Mass-density exponent from global mass-size relationships}
\label{M-R_relations}

Clumps observed in MCs are generated by turbulent flows at different scales. Nevertheless, irrespectively of the scales individual clumps originated from, their masses and sizes correlate strongly, as confirmed both by observations and numerical simulations \citep{Lars_81, Kauf_ea_10, Shet_ea_10}. That is an indication of a global mass-size relationship, with a power-law exponent $\gamma_{\rm glob}$. From observational or numerical estimates of $\gamma_{\rm glob}$, one can derive by use of equation~(\ref{gamma_x_local}) a {\it global} mass-density exponent $x(\gamma_{\rm glob})$. In our treatment, such an approach is realized in two steps: a) composition of size-mass diagrams for the set of `typical clumps' for {\it all} scales $L$ and calculation of $\gamma_{\rm glob}$ from a power-law fit of the size-mass correlation; b) derivation of $x(\gamma_{\rm glob})$. The latter quantity is representative for the whole SF process at scales $0.1\lesssim L \lesssim 30$~pc and thus can be a clue to derivation of the clump mass function. Note that even if $\gamma_{\rm glob}$ is found to vary with $L$ (i.e. no single mass-size power-law relationship is applicable), $x(\gamma_{\rm glob})$ is not expected to exhibit the same functional behavior like $x$. 

\begin{figure*} 
\begin{center}
\includegraphics[width=1.\textwidth]{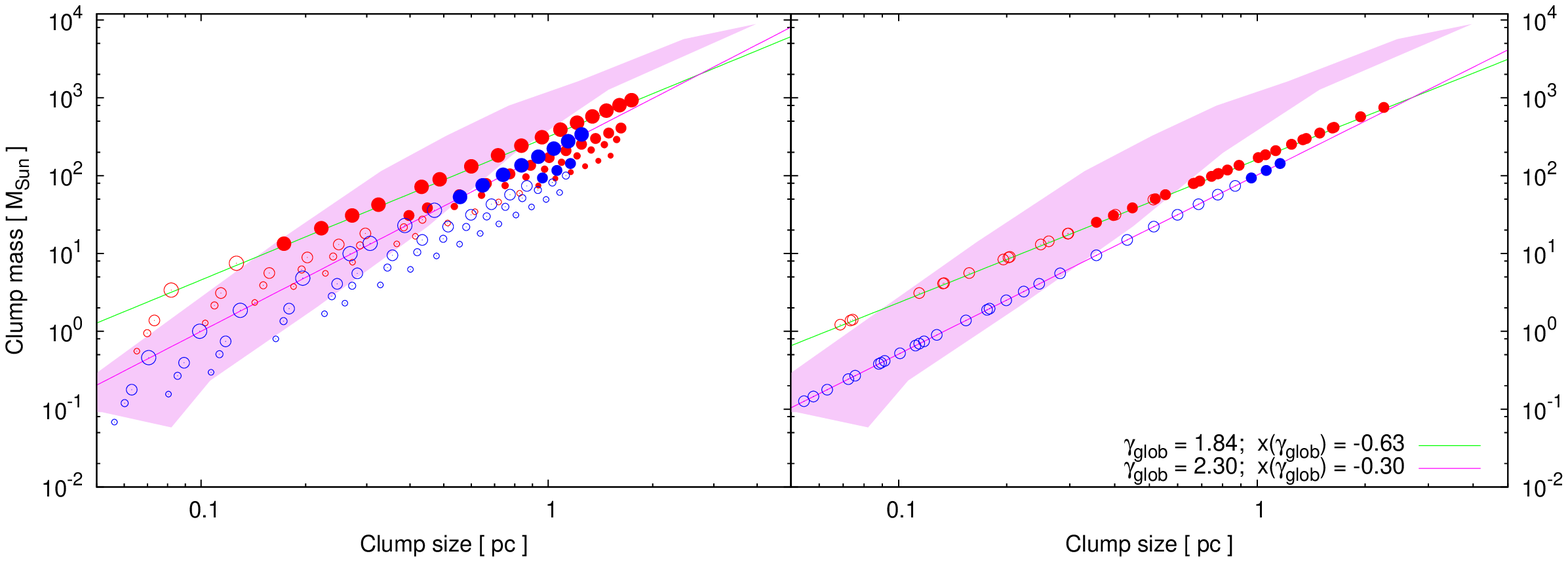}
\vspace{0.6cm} 
\caption{Global size-mass relationships, obtained from the equipartitions of gravitational vs. kinetic energy for $\beta=0.42$ (red symbols), $\beta=0.65$ (blue symbols) and adopting a scale-free $\kappa$. {\it Left:} variyng factor $f_{\rm gk}=1.0,\,1.5,\,2.0,\,4.0$ (increasing pointsize) and sets of $(b,~\kappa)$ like in the top panels of Fig.~\ref{fig_gr_kin}; {\it Right:} strongly self-gravitating clumps ($f_{\rm gk}=2.0$) and sets of $(b,~\kappa)$ like in the bottom panels of Fig.~\ref{fig_gr_kin}. Gravitationally unstable clumps are denoted by filled symbols. The zone, populated by the PPP clumps from the simulation of \citet{Shet_ea_10}, is shown (shaded area) for comparison.}
\label{fig_MR_gr_kin}
\end{center}
\end{figure*}

\begin{center}
\begin{figure} 
\includegraphics[width=80mm]{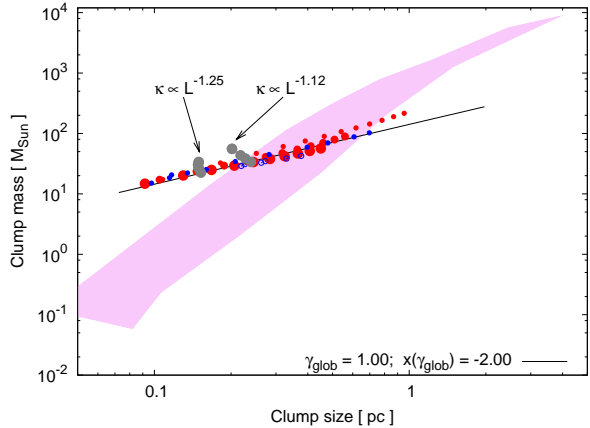}
\hspace{0cm}
\vspace{0.8cm} 
\caption{Global size-mass relationships, obtained for equipartition of the gravitational vs. kinetic and magnetic energy (equation~\ref{noth_mag}). The designations are the same like in Fig.~\ref{fig_MR_gr_kin}. The effect of varying turbulent forcing parameter $0.30\le b \le 0.70$ (increasing pointsize, $\beta=0.42$) and of adopting a scale-dependent $\kappa$ (grey symbols, $\beta=0.65$) are shown.}
\label{fig_MR_gr_noth}
\end{figure}
\end{center}

\begin{figure} 
\begin{center}
\includegraphics[width=80mm]{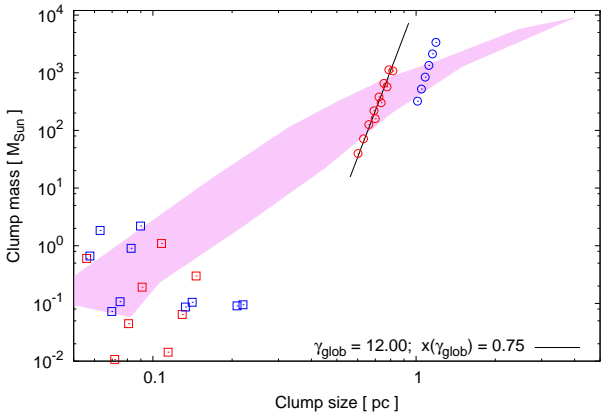}
\vspace{0.6cm} 
\caption{Global size-mass relationships, obtained from the equipartition of the kinetic vs. internal energy (squares, equation~\ref{kin_th}) and vs. magnetic energy (circles, equation~\ref{kin_mag}). In the former case, only clumps that correspond to positive $x$ are shown. The sets of $b$ and $\kappa$ are the same like in Fig.~\ref{fig_kinthmag} (top and bottom). Other designations are the same like in the previous figures.}
\label{fig_MR_kin_thmag}
\end{center}
\end{figure}

\begin{figure*} 
\begin{center}
\includegraphics[width=1.\textwidth]{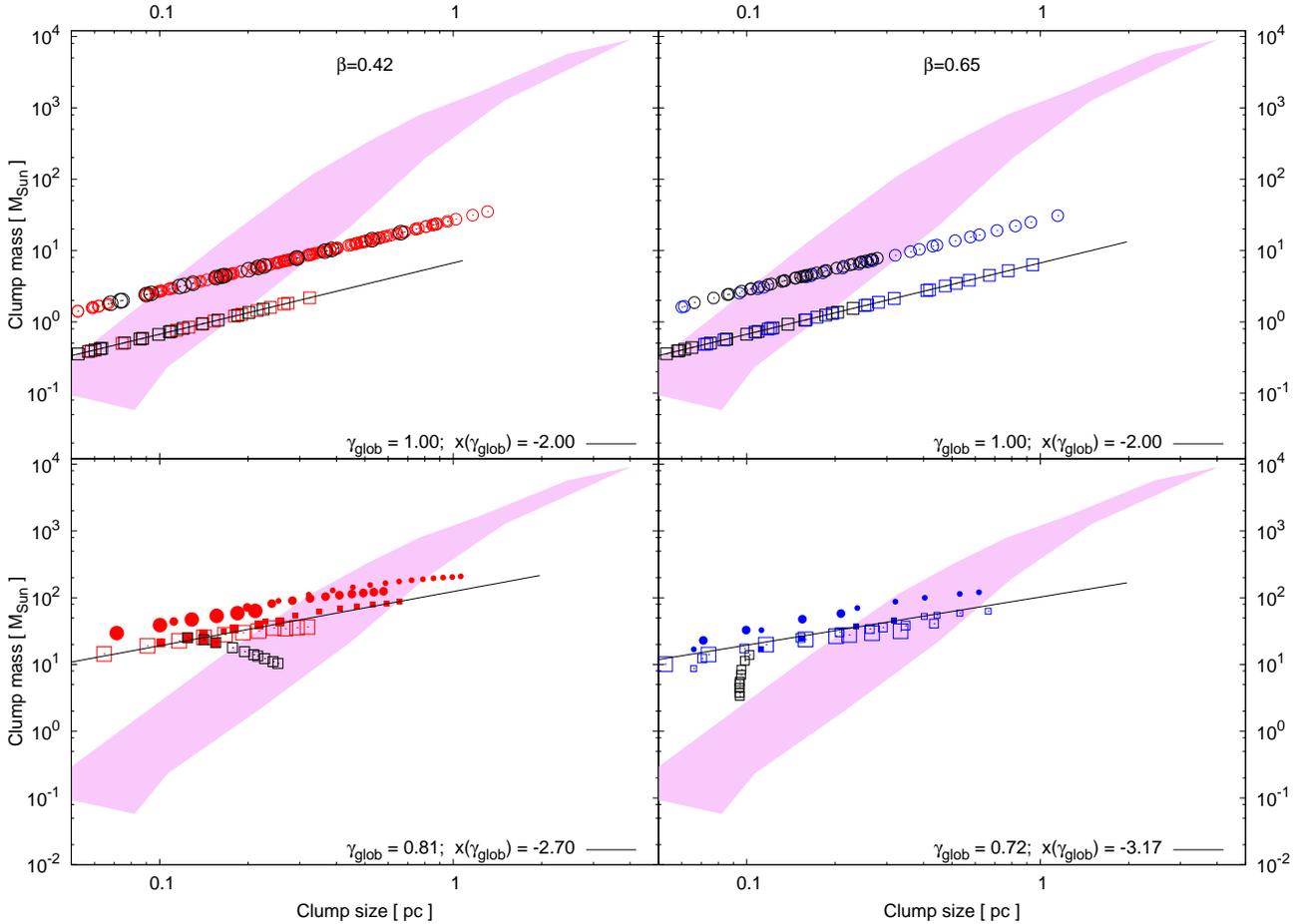}
\vspace{0.6cm} 
\caption{Global size-mass relationships, obtained from the equipartition of the gravitational vs.  internal (top) and magnetic energy (bottom). Squares denote the case $f_{\rm gt}=1$ (or, $f_{\rm gm}=1$) and circles - the case $f_{\rm gt}=2$ (or, $f_{\rm gm}=2$). Symbols with black borders are examples for solutions $\kappa\propto L^{-a}$.}
\label{fig_MR_gr_thmag}
\end{center}
\end{figure*}

Composite size-mass diagrams, obtained from the equipartition relations (equations~\ref{grav_kin}-\ref{grav_mag}), are plotted in Fig. \ref{fig_MR_gr_kin}-\ref{fig_MR_gr_thmag}, together with the results of \citet{Shet_ea_10} from simulations. The result for the equipartition of the gravitational vs. kinetic energy is illustrated in Fig. \ref{fig_MR_gr_kin}. In the left panel, we demonstrate the effect of varying the factor $f_{\rm gk}$. The larger it is, the more clump mass is required to fulfil the equipartition equation. Applying the classical Jeans criterion under the assumption of spherical clump shape, one sees that part of the clumps are gravitationally unstable (filled symbols). Most of them are generated on larger scales. There is a lower scale limit for producing unstable clumps, see e.g. Fig. 1 in \citet{VKC_11} and the comments on it. Remarkably, the points for all `typical clumps' and even average clump ensembles (the latter are not plotted for clarity) are aligned without dispersion. This result is {\it not trivial} since $x$, which determines the size-mass relationship for the `typical clump' in an individual ensemble, is a complex function of $\alpha$, $\beta$, $b$ and $\kappa$ which includes exponential and power-law terms (cf. equation~\ref{norm_units}): 
\begin{equation}
\label{clump_ml_relationship}
m_c=\frac{\pi}{6}\rho_0 l_c^3 \,L^{3x/(1-x)}\,\exp\Big(-\frac{\sigma^2}{2}\frac{2-x}{x}\Big)
\end{equation}

The slope of the obtained size-mass relationship depends exclusively on the choice of the velocity scaling parameter and is insensitive to $f_{\rm gk}$, the type of turbulent forcing and the mapping resolution. This is demonstrated in the right panel of Fig. \ref{fig_MR_gr_kin} in the case of strongly gravitating clumps as the parameters $b$ and $\kappa$ are varied simultaneously. Results for a scale-dependent $\kappa$ exhibit similar slope but essentially lower sizes and masses.

As suggested from numerous observations of dense structures in MCs, they exhibit power-law mass-size relationships of slope about 2 and larger \citep[e.g.,][]{Oni_ea_96, Tachi_ea_02}. Our `typical clumps' are expected to follow similar behaviour as they span the same range of mass and size. As seen from Fig.~\ref{fig_MR_gr_kin}, this is indeed the case with the equipartition $|W|\sim f_{\rm gk} E_{\rm kin}$. We find $\gamma_{\rm glob}\sim 2$ for the whole range of values of the velocity scaling parameter $\beta$. This result could be interpreted in terms of `typical clump' gravitational potential $\phi_c$, since in our approach for derivation of $x$ all equipartition functions (see Appendix~\ref{appendix_b}) that include gravity can be expressed through $\phi_c$ (i.e. include density to the first power). If a global size-mass relationship $m_c\propto l_c^{\gamma_{\rm glob}}$ exists, the dependence of the potential on the clump size must be
\[  \phi_c \propto m_c/l_c \propto l_c^{\gamma_{\rm glob}-1}~. \]

In view of the abovementioned studies of MCs and cloud complexes, wherein $\gamma_{\rm glob}\gtrsim 2$ was found, $\phi_c$ of the clumps should grow linearly or steeper with their size. On the other hand, $\gamma_{\rm glob}=1$ is the lower limit of observationally verifiable clump mass-size relationships.

Inclusion of a magnetic term in the equipartition by use of equation \ref{noth_mag} yields significantly a shallower slope (Fig. \ref{fig_MR_gr_noth}). It points to the critical role of the magnetic energy in the clump energy balance. This can be seen also in Fig.~\ref{fig_MR_gr_thmag} (bottom) where the results from the equipartition of the gravitational vs. magnetic energy are shown. Proper description of clump geometry is crucial in all cases with a magnetic term, as expected in view of the fact that the magnetic field is not spherically symmetric. Solutions with a scale-dependent mapping resolution ($\kappa\propto L^{-a}$) yield strange clump families and fail to reproduce the expected size-mass relationship (Fig.~\ref{fig_MR_gr_noth} and \ref{fig_MR_gr_thmag}, bottom). 

The gravitional instability of the `typical clumps' is estimated by use of the Jeans criterion, comparing $m_c$ and $m_{\rm J}(n_c)$. Excluding the ratio $n_c/n_0$ in equations \ref{ml_clumps}-\ref{nl_clumps} and replacing $m_0$ from equation~\ref{norm_units}, one obtains:
\begin{equation}
\label{mc_mj}
 m_c=\frac{\pi}{6}\rho_c l_c^3 \exp\Big(\sigma^2 \times \frac{x-1}{x} \Big)
\end{equation}

This relation holds for all members of the `average clump ensemble'. Obviously, $m_c$ depends on `typical clump' size and density in a more complex way than naively expected, due to the statistical nature of the studied objects. Thus the Jeans criterion for instability is different for each chosen equipartition.

The equipartitions of the kinetic energy (equations~\ref{kin_th} and \ref{kin_mag}) do not yield a pronounced mass-size relationship (Fig.~\ref{fig_MR_kin_thmag}). The equipartition $E_{\rm kin} \sim E_{\rm th}$ yields negative solutions $x$ within restricted spatial scale range (cf. (Fig.~\ref{fig_kinthmag}) and predicts clumps of very low sizes. On the other hand, positive solutions for $x\sim0.25 - 0.5$ are not supported by LAL10 data (Fig.~\ref{fig_kinthmag}) and correspond in general to small clumps of very low mass. However, the upper envelope of this clump population in the $m-l$ diagram apparently falls within the zone of the so called `coherent cores' ($0.04-0.1$~pc, $\sim1~M_\odot$). Indeed, this picture is to be expected from the observational definition of the coherent cores \citep{Good_ea_98} as objects wherein the contribution of the thermal velocity becomes comparable to the turbulent one, $v_{\rm turb}\sim v_{\rm th}$. The latter relation corresponds exactly to the equipartition $E_{\rm kin} \sim E_{\rm th}$ (cf. also equation~\ref{equipf_kinth}). Interestingly, the only physically meaningful solutions for the equipartition in question are obtained for the coherent cores.

The clump families, obtained from the equipartition $2 E_{\rm kin}\sim E_{\rm mag}$ exhibit anomalously steep slopes $\gamma_{\rm glob}$ that lead to values of $x\gtrsim0.75$. As evident from Fig.~\ref{fig_MR_kin_thmag}, such families could explain clumps in restricted ranges of size ($l\sim0.6-1$~pc) and mass. Those clumps are generated within restricted ranges of scales as well:  $L\gtrsim7$~pc. Equipartitions of the clump kinetic energy, without accounting for gravity, appear insufficient for description of clumps in broad ranges of mass and size. 

As already pointed out, the case of equipartition of the gravitational vs. the turbulent (kinetic) energy -- whether in strongly gravitating clumps or not -- fits well both with observations (LAL10, Fig.~\ref{fig_gr_kin}) and with simulations (\citet{Shet_ea_10}, Fig.~\ref{fig_MR_gr_kin}). That hints at the substantially lower weight of the thermal and magnetic energy in equipartitions between volume energy terms. Mass-size relationships, derived from equipartitions of $E_{\rm th}$ or $E_{\rm mag}$ with $|W|$, are instructive in that aspect (Fig.~\ref{fig_MR_gr_thmag}). The equipartition $|W|\sim f_{\rm gt} E_{\rm th}$ (equation~\ref{grav_th}) recovers perfectly the classical Jeans approach, i.e. the `Jeans-like case' in \citet{VKC_11}. Under the assumption of isothermality, $m_{\rm J}\propto n^{-1/2} \propto m^{-1/2} l^{3/2}$. If $m\propto m_{\rm J}$, than $m \propto l$, i.e. $\gamma_{\rm glob}=1$ (Fig.~\ref{fig_MR_gr_thmag}, top). In other words, the case $|W|\sim f_{\rm gt} E_{\rm th}$ produces clumps of mass, proportional to the ensemble averaged Jeans mass. Appropriate choice of the factor $f_{\rm gt}>2$ could determine the domain of gravitational instability. The equipartition $|W|\sim f_{\rm gm} E_{\rm mag}$ yields a significantly shallower slope of the mass-size relationship, similar to the case when $E_{\rm mag}$ and $E_{\rm kin}$ are both taken into consideration (Fig.~\ref{fig_MR_gr_noth}). It confirms the critical role of the magnetic energy in the balance between clump volume energy terms. We point out again the significance of the clump mapping in this case, the choice $\kappa \propto L^{-a}$ leads to unrealistic results. We attribute them to the specific clump geometry shaped by the presence of magnetic fields.

\section{Summary and conclusions}    \label{Summary}
This work is an attempt for theoretical substantiation of the clump mass-density relationship $n \propto m^x$ and for a plausible estimation of its exponent $x$. Due to the turbulence and the dynamics of the ISM, such a relationship has a statistical significance and therefore a statistical approach was chosen. The time frame of our consideration refers to the transition between two main stages of the MC evolution: the predominantly turbulent epoch and the period of subsequent hierarchical gravitational collapse \citep{BP_ea_11}. At that timescale, condensations in the initially diffuse, atomic ISM become molecular, gravitationally bound and magnetically supercritical \citep{VS_10}. The turbulence in the newly formed MCs becomes essentially supersonic, due to the drastic decrease of temperature \citep{VS_ea_07, Baner_ea_09}. The turbulent inertial range covers spatial scales from several tenths to several decades of parsec. We assume steady state and validity of the ergodic hypothesis. Using the scaling laws for velocity, density and magnetic field and equipartition relations for various forms of energy for a statistically averaged (`typical') clump at different scales $L$, we obtain equations to derive the mass-density exponent $x$.

The obtained results could be summarized as follows.

1. Generally, the solutions $x$ correspond, both as range of values and as functional behavior, to the (sub)structure of MCs, recently studied by LAL10 by use of extinction maps. However, this agreement depends substantially on the chosen equipartition relation. The general conclusion is twofold: i) equipartitions which do not include gravitational energy are not relevant for derivation of $x$, and, ii) in contrast to gravitational and kinetic energy, magnetic and thermal energy do not play significant role for the clump mass-density relationship.
 
Good fits to the observational data of LAL10 are obtained by variations of the turbulent forcing parameter $b$ and the mapping resolution $\kappa$ and are not sensitive to the choice of velocity scaling law (the value of $\beta$). That suggests a connection between those parameters and the specific structure and evolution of the clouds and, possibly, to an universally applicable range of variations of $\beta$. However, to model the cloud structure in more details, one needs additional data concerning the evolutionary status of the cloud and physical conditions therein. Our aim in this work is simply to demonstrate the capacity of the proposed statistical approach, combined with equipartition equations to derive $x$. Finally we point out that variations of the clump parameters within statistically fiducial ranges (`average clump ensemble') also cover the diversity of MC structure, exhibited by the observational analysis of LAL10.

2. The composite mass-size diagrams, constructed for all `typical clumps', generated at different scales, yield a global value of mass-size exponent $\gamma_{\rm glob}$ and hence a global mass-density exponent $x(\gamma_{\rm glob})$. The latter quantities are representative for the diversity of Galactic MCs and of their substructures. An immediate conclusion from the clump mass-size diagrams is that an equipartition relation between the clump gravitational and kinetic energy provides realistic estimates both of $x$ and of $x(\gamma_{\rm glob})$. This lends additional support to our previous conclusion about the secondary importance of the clump magnetic and thermal energy for the clump mass-density relationship. On the other hand, the implied geometry in the description of magnetic field is crucial when a magnetic term is present in the equipartition relations. 

3. In contrast to the local value $x$, the behaviour of the global mass-density exponent $x(\gamma_{\rm glob})$, derived from different equipartitions which include gravitational and kinetic energy (Fig.~\ref{fig_MR_gr_kin}), does not depend on the turbulent forcing, on the mapping resolution and on the equipartition coefficient. It leads us to conclude that the obtained estimates of the global slopes of the mass-size and mass-density relationships are universal for the variety of Galactic MCs and their substructures. However, those slopes depend notably on the choice of the velocity scaling law. This finding could be interpreted in terms of the prehistory of MC formation. Turbulence in the MC is inherited from the natal diffuse ISM and hence its power spectrum is preliminarily determined. We suppose that the prehistory of MC formation affects the clump mass function as well, through the value of $\gamma_{\rm glob}$.
 
4. The analysis of our results demonstrates the importance of the different forms of energy in the dynamics of gas clumps on the verge between the purely turbulent epoch in the MC evolution and the stage of hierarchical gravitational collapse. The key factors for clump formation and structure are gravity and turbulence. Our assumption of spherical symmetry of the clumps implicitly diminishes the significance of the magnetic field. The role of the thermal energy would increase if spatial scaling of the temperature is included in the description. 

Possible future extension of this work is to implement spatial scaling of the gas temperature and to improve the geometrical description of clumps, retaining our basic physical assumptions. The estimates of the global clump mass-size slope could be used to develop a model of the clump mass function as a key to the stellar initial mass function. It would be also interesting to extend the time frame of our consideration, including earlier (non-gravitational) and later (gravitationally dominated) stages of the MC evolution.  

{\it Acknowledgement:} T.V. acknowledges support by the {\em Deutsche Forschungsgemeinschaft} (DFG) under grant KL 1358/13-1 and by the Scientific Research Foundation, Ministry of Education and Sciences, Bulgaria, under contract VU-F-201/06. S. D. wants to thank Ivan Stefanov from the Technical University in Sofia for the stimulating discussions on the theoretical problems considered in the present work. We are grateful to Christoph Federrath, Rahul Shetty and Robi Banerjee for their comments and suggestions in the process of preparation of this paper. 


\label{lastpage}

\newpage
\appendix
\section{Relations between the normalization units}
\label{appendix_a}
If $N_ l$ is the number of clumps of size $l$ and $s_l\equiv\ln(l/l_0)$, the requirement of volume conservation gives:
\begin{eqnarray}
 V  = \sum\limits_l a l^3 N_l=N \sum\limits_l a l^3 (N_l/N) = N\sum\limits_l a l^3 p(l)
\end{eqnarray}
where $p(l)=p(s_l)\,ds_l$ is the probability (fraction) of clumps of size $l$ and $N$ is the total number of clumps, generated at $L$. Thus, since $p(s_l)\,ds_l=p(s)\,ds$ (cf. equations \ref{param_x}-14), one obtains
\begin{eqnarray}
 V & \simeq & aN\int\limits_{-\infty}^{+\infty} l^3 p(s_l)\,ds_l= aNl_0^3\int\limits_{-\infty}^{+\infty}\Big(\frac{l}{l_0}\Big)^3 p(s)\,ds \nonumber
\end{eqnarray}
Using equation \ref{eq_n-l}, one gets:
\begin{eqnarray}
 V & = & \frac{aNl_0^3}{\sqrt{2\pi\sigma^2}}\int\limits_{-\infty}^{+\infty}\Big(\frac{n}{n_0}\Big)^{\frac{1-x}{x}}  \exp\Bigg[-\frac{1}{2}\bigg( \frac{s -s_{\rm max}}{\sigma}\bigg)^2\Bigg]ds \nonumber
\end{eqnarray}
Further, from $n/n_0=\exp(s)$ and by rewriting the integral for variable $y\equiv (s -s_{\rm max})/\sqrt{2\sigma^2}$, we obtain:
\begin{eqnarray}
 V = \frac{aNl_0^3}{\sqrt{\pi}}\exp\Big({\frac{1-x}{x}s_{\rm max}}\Big)\int\limits_{-\infty}^{+\infty} \exp(cy-y^2) dy~~, \nonumber\\
c\equiv \sqrt{2\sigma^2}\frac{1-x}{x}
\end{eqnarray}
The right-hand side of this equation contains a Gauss integral with an analytical solution. Taking into account that $V=aL^3$, we derive finally equation \ref{volume_conservation}.

The requirement for mass conservation yields equation \ref{mass_conservation} in analogical way, starting from:
\begin{equation}
\label{derive_mass_conserv}
 M\simeq N\int\limits_{-\infty}^{+\infty} m p(s_m)\,d s_m=N m_0\int\limits_{-\infty}^{+\infty} \Big(\frac{n}{n_0}\Big)^{\frac{1}{x}} p(s)\,ds
\end{equation}

\section{Equipartition functions}\label{appendix_b}
These functions are used to derive the clump mass-density exponent $x$ at a given scale. They are obtained from the equipartition relations, given in Section~\ref{energy_equipartitions}, through the following procedure:
\begin{itemize}
 \item[-] Expressions for different energies (\ref{W_clumps} - \ref{ek_clump}) are substituted in the equipartition relations (\ref{grav_kin}) - (\ref{grav_mag})
 \item[-] The quantities $l_c$ and $m_c$ are excluded from the equations via substitution with $\rho_c$ by use of equations~(\ref{ml_clumps}) and (\ref{nl_clumps}).
 \item[-] The `typical clump' density $\rho_c$ is expressed through equation~(\ref{eq_nc}).
 \item[-] Clump mass normalization unit $m_0$ is expressed through $\rho_0$ and $l_0$ (equation~\ref{norm_units}).
\end{itemize}
The obtained equipartition functions are listed below. 

\subsection{Gravitational vs. kinetic energy}
From equation~(\ref{grav_kin}):
\begin{eqnarray}
\label{equipf_grav_kin}
 Q_{\rm gk}(x) & = & \frac{\pi}{5}z_c G\rho_0 l_0^2 \exp\bigg[-\sigma^2\times\Big(\frac{4x+2}{6x}-\frac{x-1}{x}\Big)\bigg] \nonumber\\ 
& & -f_{\rm gk}\frac{u_0^2}{2} \Big( \frac{l_0}{4~\rm pc}\Big)^{2\beta}\!\!\exp\Big(-\sigma^2\times\frac{2\beta+(3-2\beta)x}{6x}\Big)
\end{eqnarray}

\subsection{Gravitational vs. kinetic and magnetic energy}
From equation~(\ref{noth_mag}):
\begin{eqnarray}
\label{equipf_nothmag}
 Q_{\rm gkmag}(x) & = & \frac{\pi}{5}z_c G\rho_0 l_0^2 \exp\bigg[-\sigma^2\times\Big(\frac{4x+2}{6x}-\frac{x-1}{x}\Big)\bigg] \nonumber\\
 & & -u_0^2 \Big( \frac{l_0}{4~\rm pc}\Big)^{2\beta}\exp\Big(-\sigma^2\times\frac{2\beta+(3-2\beta)x}{6x}\Big) \nonumber\\
 & & - \frac{B^2}{8\pi\rho_0}\exp\Big(-\frac{\sigma^2}{2}\Big)
\end{eqnarray}


\subsection{Kinetic vs. thermal or magnetic energy}
From equations~(\ref{kin_th}) and (\ref{kin_mag}): 
\begin{eqnarray}
\label{equipf_kinth}
Q_{\rm kth}(x) & = & u_0^2 \Big( \frac{l_0}{4~\rm pc}\Big)^{2\beta} \exp\Big(-\sigma^2\times\frac{2\beta+(3-2\beta)x}{6x}\Big)\nonumber\\
 & & -3\exp\Big(-\frac{\sigma^2}{2}\Big)\frac{\Re T}{\mu} 
\end{eqnarray}
\begin{eqnarray}
\label{equipf_kinmag}
Q_{\rm kmag}(x) & = & u_0^2 \Big( \frac{l_0}{4~\rm pc}\Big)^{2\beta} \exp\Big(-\sigma^2\times\frac{2\beta+(3-2\beta)x}{6x}\Big) \nonumber\\
 & & - \frac{B^2}{8\pi\rho_0}\exp\Big(-\frac{\sigma^2}{2}\Big)
\end{eqnarray}

\subsection{Gravitational vs. thermal or magnetic energy}
From equations~(\ref{grav_th}) and (\ref{grav_mag}): 
\begin{eqnarray}
\label{equipf_grav_th}
 Q_{\rm gth}(x) & = & \frac{\pi}{5}z_c G\rho_0 l_0^2 \exp\bigg[-\sigma^2\times\Big(\frac{4x+2}{6x}-\frac{x-1}{x}\Big)\bigg] \nonumber\\
 & & -\frac{3}{2}f_{\rm gt}\exp\Big(-\frac{\sigma^2}{2}\Big)\frac{\Re T}{\mu} 
\end{eqnarray}
\begin{eqnarray}
\label{equipf_grav_mag}
 Q_{\rm gmag}(x) &=& \frac{\pi}{5}z_c G\rho_0 l_0^2 \exp\bigg[-\sigma^2\times\Big(\frac{4x+2}{6x}-\frac{x-1}{x}\Big)\bigg] \nonumber\\
 & & - f_{\rm gm}\frac{B^2}{8\pi\rho_0}\exp\Big(-\frac{\sigma^2}{2}\Big)
\end{eqnarray}

\end{document}